\documentclass[epj]{svjour}
\usepackage{graphics}
\usepackage{amsmath}
\begin{document}
\title{Strangeness production time and the $K^+/\pi^+$ horn}
\author{Boris Tom\'a\v{s}ik\inst{1}\inst{2} \and Evgeni E Kolomeitsev\inst{3}
}                     
%
%
\institute{Fakulta pr\'irodn\'ych vied, Univerzita Mateja Bela,
Tajovsk\'eho 40, 97401 Bansk\'a Bystrica, Slovakia \and \'Ustav
vedy a v\'yskumu, Univerzita Mateja Bela, Cesta na amfite\'ater 1,
97401 Bansk\'a Bystrica, Slovakia \and University of Minnesota,
School of Physics and Astronomy, 116 Church Street SE, Minneapolis,
55455 Minnesota, USA}
\date{\today}
%
\abstract{ %
We construct a hadronic kinetic model which describes production 
of strange particles in ultrarelativistic nuclear collisions in the
energy domain of SPS. We test this model on description of the sharp
peak in the excitation function of multiplicity ratio $K^+/\pi^+$ and 
demonstrate that hadronic model reproduces these data rather well. 
The model thus must be tested on other types of data in order to 
verify the hypothesis that deconfinement sets in at lowest SPS energies. 
} 
\maketitle
\section{Introduction}
\label{intro}

The central issue in studies with ultrarelativistic heavy ion
collisions is to map the phase diagram of strongly interacting
matter. From lattice QCD and numerous investigations with
effective models we know that at high enough energy density
hadronic matter melts to a phase where quarks and gluons are the
relevant degrees of freedom. Data on jet quenching at RHIC
energies indicate very clearly that such a deconfined phase has
been produced there \cite{brahmsWP,phobosWP,starWP,phenixWP}. This
does not, however, answer the question {\em where is the
threshold} for deconfinement.

In order to find the minimum collision energy at which the
hadronic description of collision dynamics turns inadequate, a
natural choice is to study various excitation functions, i.e.\
dependences of quantities on the collision energy. The expectation
here is that a change of the quality of the system would
demonstrate itself as a non-monotonic dependence of some
excitation function. A set of interesting excitation functions has
been observed in the energy scan at the SPS. These include: i) a
sharp peak at projectile energy 30 $A$GeV of the ratio of
positive kaon to pion multiplicities (``the horn''); ii) a change in
the slope of number of produced pions per participating nucleon
(``the kink''); and
iii) a plateau in the excitation function of kaon mean $p_t$ which
is otherwise increasing function of the collision energy (``the
step'') \cite{na49data,lung}. In addition to
this, event-by-event fluctuation of the $K/\pi$ multiplicity
ratio grows when the collision energy is lowered and reaches 
maximum at lowest SPS energies, though we have to note that there
are no data at lower energies from AGS.

These observed features possibly indicate qualitative changes
of the system. A possible explanation is that they are caused by
the onset of deconfinement. Such a {\em positive} reasoning is
insufficient, however, for claims of this kind of discovery. In
fact, one would have to demonstrate that any purely hadronic
description of the data fails. In this paper we embark on such a
programme and address one of the excitation functions: the horn.

Currently, the data have been addressed in many different
approaches. Transport codes generally fail in reproducing some of
the observed multiplicities as functions of collision energy and
thus fail in ratios \cite{transport}. A three-fluid hydrodynamic
simulation does not correctly reproduce the multiplicity of
negative pions \cite{toneev}. Statistical hadronisation model
leads to a broad peak of the $K^+/\pi^+$ ratio as a function of
collision energy which is put in connection with transition
from baryon-dominated to meson-dominated entropy \cite{statmodel}.
The peak is much broader than observed, however. Data are better
reproduced in a statistical hadronisation fit in which strange
species are suppressed with respect to chemical equilibrium by a
stran\-ge\-ness suppression factor \cite{becat}. A question appears
then, what is the mechanism leading to the particular value of
stran\-ge\-ness suppression factor?

The data are successfully interpreted in framework of  the
so-called Statistical Model of the Early Stage (SMES)
\cite{GazGor}. The model predicts first order phase transition and
the existence of mixed phase, which sets in about the beam energy
of 30 $A$GeV. Another important ingredient of the model is the
assumption of an immediate chemical equilibration of the
primordial production of quanta. A non-equilibrium kinetic
calculation reproducing the data which also includes first order
phase transition has been proposed recently \cite{kincal}.

We shall construct hadronic non-equilibrium kinetic model and
try to reproduce the peak in the ratio of multiplicities
$K^+/\pi^+$. It will be shown that these data can be interpreted
in framework of such a model and therefore hadronic description
must be tested on other types of data until it can be safely
ruled out.

\begin{figure*}[ht]
\centerline{\resizebox{0.8\textwidth}{!}{\includegraphics{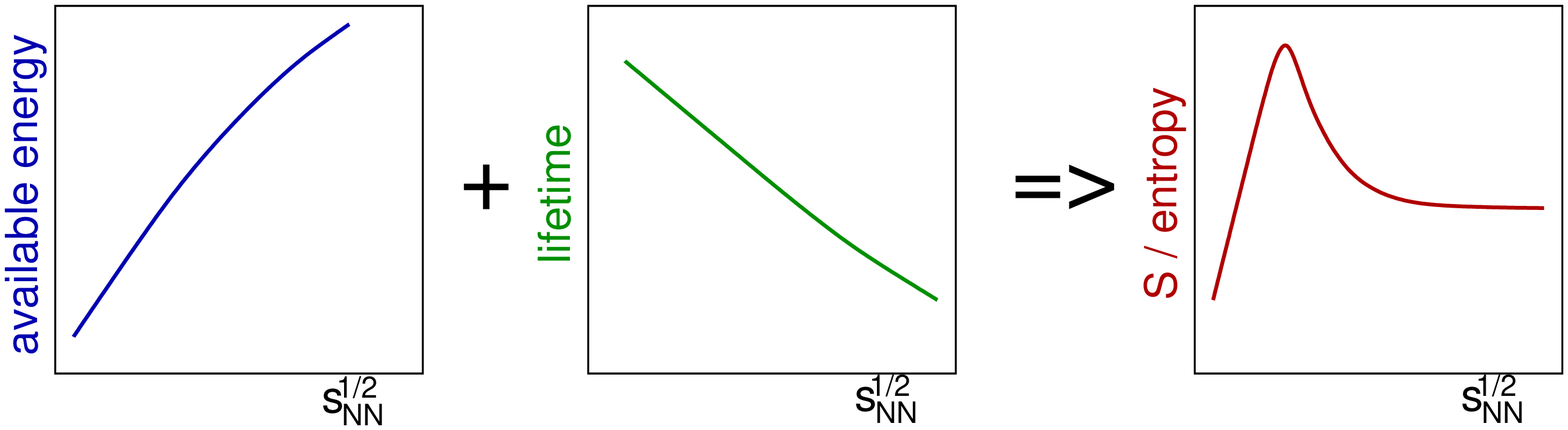}}}
\caption{(colour online)
If energy available for production of strange particles
increases with collision energy while the lifetime of fireball decreases
with increasing collision energy, these two excitation functions might
combine into a peak in strangeness-to-entropy ratio as a function of
collision energy. The ratio of strangeness to entropy is experimentally
accessible through the ratio of positive kaon to pion multiplicity.}
\label{fig-cartoon}
\end{figure*}


\section{Strangeness production: cartoon description}
\label{cartoon}

Strange particles are produced in nucleon-nucleon reactions,
though in smaller fraction to non-strange ones than in nuclear
collisions. We shall assume that the corresponding surplus of
strange particles is produced in secondary reactions of hadrons
generated in nuclear collisions. Then, two aspects are important
for strangeness production: available energy and time.

The available energy is microscopically represented\linebreak 
through energy
density. At typical momenta of particles in the considered 
environment, many
secondary reactions run close to the threshold of kaon and hyperon
production. In such a case, production rates depend strongly on
incoming momenta and densities of reactants, which both depend on
temperature (if it is defined). Higher energy density (and
temperature) implies higher production rates.

Since strangeness is under-represented with respect to chemical 
equilibrium in nuc\-le\-on-nuc\-le\-on collisions, its\linebreak
amount in the
system will increase with time until it saturates. Thus longer
total lifespan of the system would also mean larger relative
amount of strange particles. This picture becomes slightly more
complicated in an expanding and cooling system, but the simple
assumption that the amount of strange particles grows with time is
good enough for the cartoon-like explanation offered in this
section.

Now we can imagine the following explanation of the $K^+/\pi^+$
horn (Fig.~\ref{fig-cartoon}): the energy density within the
system which is available for production of
strange particles is an increasing function of the collision
energy. We could also assume that the total lifespan of the
fireball decreases as the collision energy grows. These two
features might be combined in such a way, that the
strangeness-to-entropy ratio---which is basically measured by the
$K^+/\pi^+$ ratio---shows a sharp peak as it is indeed observed. In the
next Section we shall describe this model in more technical terms.

\begin{figure}[t]
\centerline{\resizebox{0.4\textwidth}{!}{\includegraphics{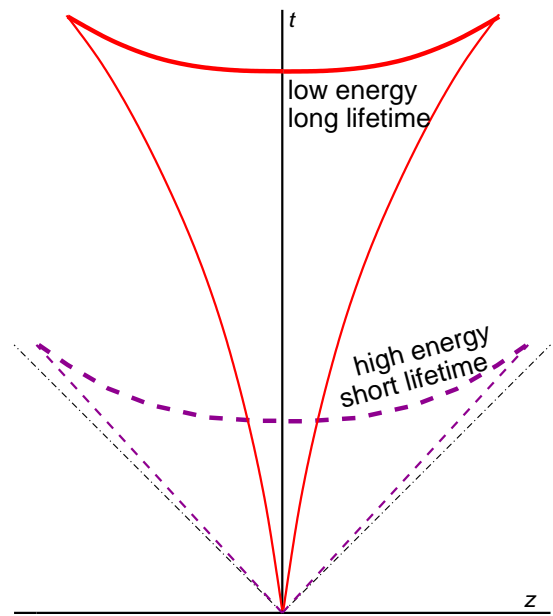}}}
\caption{(colour online)
Evolution scenarios and freeze-out hypersurfaces in a space-time
diagram for fireballs created at lower collision energy with strong
stopping and re-expansion (solid line), and higher collision energy in
nuclear transparency regime (dashed line). Longer lifetime of a fireball
does not necessarily imply larger extent in longitudinal direction.}
\label{fig-accel}
\end{figure}
The assumed dependence of the total lifespan on collision energy
may result from an interplay between stopping and subsequent
accelerated expansion. At low collision energy we expect strong
slowing down of the incident nucleons. Then the system needs time
in order to build up expansion from pressure and to expand up to
the breakup density. At higher collision energy stopping is weaker
and the incident nucleons continue longitudinal movement. Thus
expansion is present from the first moments of fireball evolution
and it takes shorter time to get to the critical breakup density.
Let us also stress that the assumption of shorter lifespan at
higher collision energy does {\em not} clearly contradict any
existing data. It does {\em not} imply larger longitudinal extent
of the fireball and a larger freeze out volume
(Fig.~\ref{fig-accel}). Hence, it also does not necessarily lead
to higher multiplicity at lower collision energy. In fact, the
correspondence between lifetime and the longitudinal size is
model-dependent and one usually thinks about Bjorken
boost-invariant expansion \cite{bjork} when making such a
connection, but this is not the generally applicable hydrodynamic 
solution.


\section{The kinetic model}
\label{kinmod}

We shall construct a kinetic model for the production of strange
particles. Such a calculation requires knowledge of the local densities of
reacting species. This is naturally provided in so-called
hydro-kinetic or flavour kinetic models 
\cite{kora,komura,kame,knoll,knoll2,brko}.
It is known, however, that most hydrodynamic and kinetic models
have problems with reproducing the space-time extent of the
fireball measured by femtoscopy. Therefore, we shall not use
such a model in order to calculate the space-time
evolution of the fireball. Instead, we will parametrise the
evolution of the density. In this way, we do not couple our
approach directly to the microscopic structure of the matter or
to its equation of state. On the other hand, we can choose many
different parametrisations of the fireball evolution and check
which of them might lead to results consistent with data. Our
ansatzes for these parametrisation  will be constructed with one eye
on femtoscopy data and hadronic spectra.

As we are now only interested in {\em ratios} of multiplicities
it will be sufficient to calculate the {\em densities} of individual
species in the kinetic approach. For densities of kaons with positive
strangeness we derive
\begin{equation}
\frac{d\rho_K}{d\tau} = \frac{d\,}{d\tau}\, \frac{N_K}{V} =
- \frac{N_K}{V}\, \frac{1}{V}\, \frac{dV}{d\tau}
+ \frac{1}{V}\, \frac{dN_K}{d\tau}\, .
\end{equation}
Notice that the first term on the right hand side $1/V\, dV/d\tau$ 
actually includes the expansion rate which will
follow from our assumption for density evolution. The second term
includes the rate of change of the number of kaons. It can be
divided into two contributions: production rate and annihilation
rate. The former stands for all processes which produce a kaon
while the latter includes processes which destroy kaons. They are
determined from cross-sections of these processes and evolved
densities. Thus the master equation for kaon density reads
\begin{equation}
\label{maseq}
\frac{d\rho_K}{d\tau} = \rho_K \,
\left ( - \frac{1}{V}\, \frac{dV}{d\tau}\right )
+ {\cal R}_{\rm gain} - {\cal R}_{\rm loss}\, .
\end{equation}

\subsection{Production and annihilation}
\label{proan}

Densities of $K^+$, $K^0$, $K^{*+}$, and $K^{*0}$ are evolved
according to eq.~\eqref{maseq}. Here we always assume that all
particles keep their vacuum properties.
The gain term has in our treatment basically two types of contributions
\begin{equation}
{\cal R}_{\rm gain} = \sum_{i\, j\, X} \langle v_{ij}\sigma_{ij}^{KX}\rangle
\frac{\rho_i\rho_j}{1+\delta_{ij}} + \rho_{K^*} \Gamma_{K^*}\, ,
\end{equation}
where the sum goes over two-to-N processes leading to production
of kaons and the second term is for $K^*$ decay into $K$. The term
in angular brackets is cross-section of a process multiplied with
{\em relative} velocity of the reacting species and averaged over
the distribution of relative velocities. For the sake of this
averaging we always assume thermal distribution of velocities. The
annihilation term is obtained in a similar way
\begin{equation}
{\cal R}_{\rm loss} = \sum_{i\, X} \langle v_{Ki} \sigma_{Ki}^X\rangle
\frac{\rho_K\rho_i}{1+\delta_{Ki}}\, .
\end{equation}

In a real calculation, it is impossible to include all processes that
create or destroy a kaon. Only the most important ones are included
here:
\begin{itemize}
\item Associated production of kaon and a hyperon in reactions of
$\pi N \leftrightarrow KY$ and $\pi \Delta \leftrightarrow KY$. In order
to keep the detailed balance, these reactions are included in both
directions, i.e., creating and annihilating kaons.
\item Meson-meson reactions of $\pi\pi$, $\pi\rho$, and $\rho\rho$ are
also included in both directions.
\item $K^*$ production from $\pi K$ collisions and its decay.
\item Reactions of $\pi Y \leftrightarrow K\Xi$.
\item Baryon-baryon reactions which lead to kaons in the final state
are included only in the gain term.
\end{itemize}
For details of parametrisations of all the used cross-sections
we refer the reader to \cite{ktopi}. Note that no reactions
involving antibaryons have been taken into account. Since nuclear collisions
create a {\em baryon rich} environment the error due to this
simplification is reasonable. It is largest at the highest SPS
energy (we do not apply this model to higher energies) where we
estimate it around 10\%.

Other species than kaons with $S>0$ are not treated explicitly in
the kinetic approach. Chemical reactions changing their numbers
are swift and therefore we can assume that non-strange species are
kept in chemical equilibrium during the whole evolution. For
species with $S<0$---which include $K^-$ and $\bar K^0$ as well
as $\Lambda$, $\Sigma$, $\Xi$, and $\Omega$---we assume that they
are in {\em relative} chemical equilibrium, i.e., that all the
strange quarks are distributed in the system according to
principle of maximum entropy although their total number is given
by the number of produced strange antiquarks contained in kaons.

\subsection{The ansatz for expansion}
\label{expan}

We shall assume that the evolution of fireball densities
consists basically from two periods: initial accelerating period
and later scaling expansion with power-law dependence of density
on time. Since baryon number is a conserved quantum number,
evolution of baryon density actually determines the expansion of
the volume. We write it as
\begin{equation}
\label{npar}
\rho_Q(\tau) = \left \{
\begin{array}{lcl}
\rho_{Q0} (1 - a\tau - b\tau^2)^\delta & : & \tau < \tau_s \\
\frac{\rho_{Q0}^\prime}{(\tau - \tau_0)^\alpha} & : & \tau \ge
\tau_s
\end{array} \right . \, ,
\end{equation}
where $\rho_{Q0}$, $a$, $b$, $\rho_{Q0}^\prime$, $\tau_0$,
$\tau_s$, $\alpha$ and $\delta$ are parameters which can be tuned.
They can be put in relation to the total lifetime, initial
density, initial density decrease etc.\ \cite{ktopi}. The
subscript $Q$ stands for any conserved quantum number; in our
treatment we explicitly take care of baryon number and the third
component of isospin. This parametrisation is motivated by
simplicity: at the beginning the simplest way of encoding the
acceleration of expansion is through second order polynomial,
while the power law in the end is commonly used in analysis of
correlation radii.

The energy density is parametrised in a similar way
\begin{equation}
\label{epar}
\varepsilon(\tau) = \left \{
\begin{array}{lcl}
\varepsilon_{0} (1 - a\tau - b\tau^2) & : & \tau < \tau_s
\\ \frac{\varepsilon_{0}^\prime}{(\tau - \tau_0)^{\alpha/\delta}} & : & \tau \ge
\tau_s
\end{array} \right . \, ,
\end{equation}
so the parameter $\delta$ represents a simple way of choosing the
equation of state.

By exploring a range of parameters we can tune our parametrisation
between the well known Bjorken \cite{bjork} and Landau
\cite{landau} hydrodynamic solutions of fireball expansion.


\subsection{Initial and final conditions}
\label{ific}

Since kaons are also produced in nucleon-nucleon interactions, 
there must be some initial kaon density due to primordial kaon
production in collisions of incident nucleons. It is estimated
\cite{ktopi} from a compilation of kaon production data in
nucleon-nucleon collisions \cite{garo}.

Then, species containing strange quarks must balance the strange
antiquarks such that the total strangeness vanishes. Among
themselves they are in relative equilibrium.

The most important choice is that of the total lifetime of the
fireball and the initial energy density. This, together with the 
final state densities, basically determines the whole evolution 
scenario.

The energy density and number densities in the final state, i.e.,
at the end of parametrisations \eqref{npar} and \eqref{epar}, are
chosen so that they correspond to the measured values. The values
from experiment were calculated from results of chemical freeze
out fit with a statistical model \cite{becat}.

Thus the construction of our model guarantees that we end up with
the correct final state values of energy density, baryon density,
and the density of third component of isospin. It still does not
guarantee, however, the correct $K^+$ abundance, as kaons are
produced kinetically. But once we get the right amount of kaons,
together with the densities this implies the correct value of
temperature and chemical potentials. Consequently, the whole
chemical composition is correct in such a case.


\subsection{Kaon and $\Lambda$ production}
\label{kalam}

From what has been said it should be now easy to understand how
the production of kaons, antikaons, and lambdas is steered by
different knobs. This is important as we shall fit the ratios of
their multiplicities to pion multiplicity. 
The amount of kaons is mainly given by the
lifetime of the whole system and also by the initial energy density.
The strange quarks are distributed---in a simplified
picture for the sake of this explanation---among antikaons and
hyperons. The distribution of these quarks among
antikaons and hyperons is given in  relative chemical equilibrium
by the temperature and baryonic chemical potential.

In summary: lifetime determines the amount of kaons, and
subsequently temperature specifies how many an\-ti\-ka\-ons and hyperons
there will be.


\section{Results}
\label{res}

We have run our hadronic kinetic model for various sets of model
parameters for Au+Au collisions at projectile energy 11.6 $A$GeV
(highest AGS energy) and then Pb+Pb collisions at 30, 40, 80, and
158 $A$GeV (SPS), so that we safely cover the region where the
$K/\pi$ horn appears. For each value of initial energy density and
the total lifetime we evolve  kaon densities. At the end, we
add the feed-down to kaons and pions due to resonance decays. (Details
of this procedure can be found in \cite{ktopi}.)

\begin{figure}[t]
\centerline{\resizebox{0.47\textwidth}{!}{\includegraphics{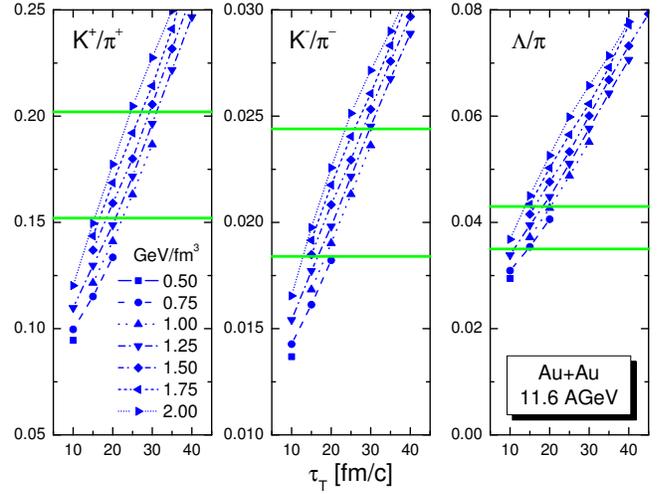}}}
\caption{(colour online)
The ratios of multiplicities 
$K^+/\pi^+$ (left panel), $K^-/\pi^-$ (middle), and
$\Lambda/\pi$ (right) calculated in scenarios with various initial
energy densities and total lifetimes of the fireball for Au+Au collisions
at projectile energy 11.6~$A$GeV. The plots show dependence of the
ratios on the {\em total} lifetime of the fireball. Different
curves correspond to different initial energy densities. Horizontal
lines indicate the 1$\sigma$ intervals around the measured values.
}
\label{f:AGS}
\end{figure}

\begin{figure}[]
\centerline{\resizebox{0.47\textwidth}{!}{\includegraphics{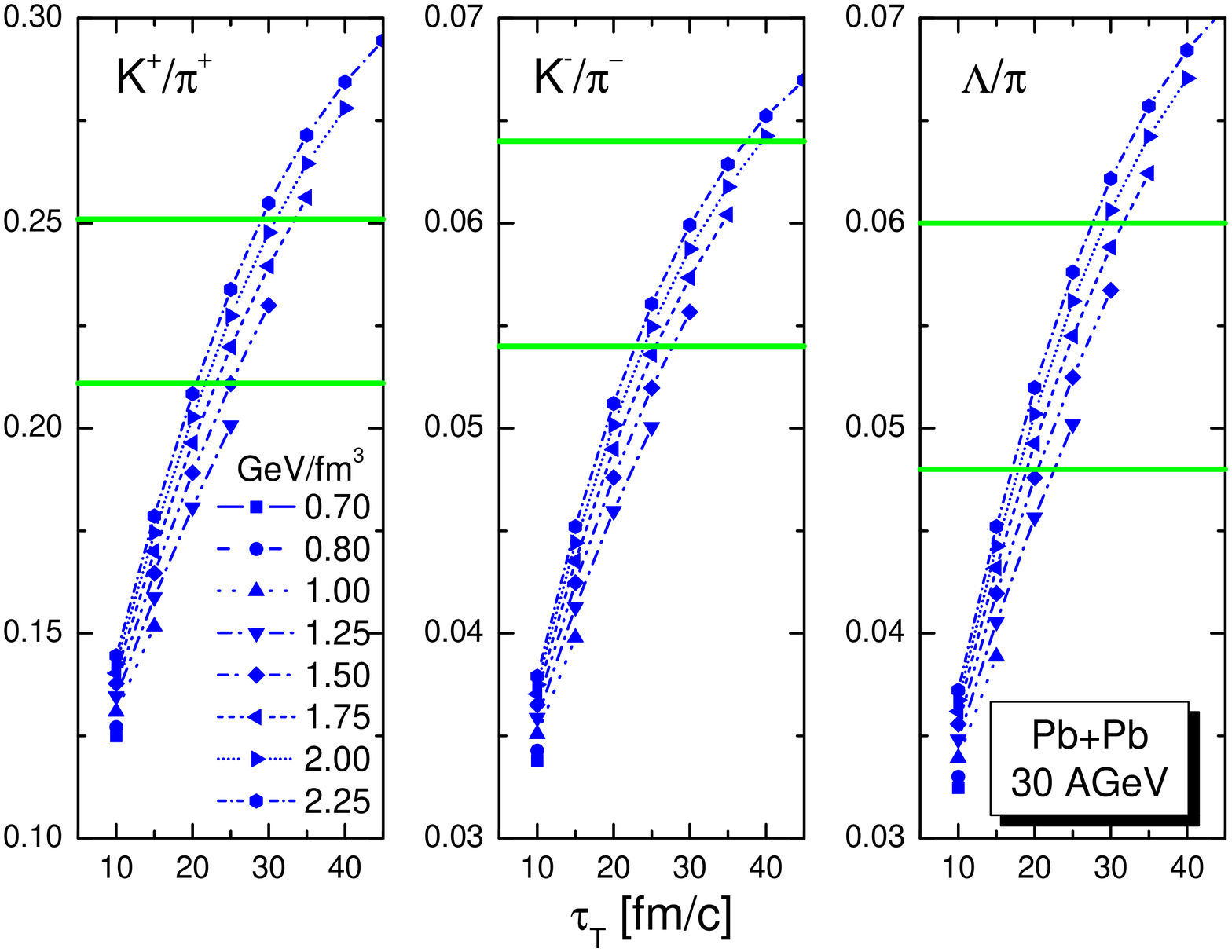}}}
\caption{(colour online) Same as Fig.~\ref{f:AGS} but for Pb+Pb
collisions at projectile energy 30~$A$GeV.
}
\label{f:tri}
\end{figure}

\begin{figure}[t]
\centerline{\resizebox{0.47\textwidth}{!}{\includegraphics{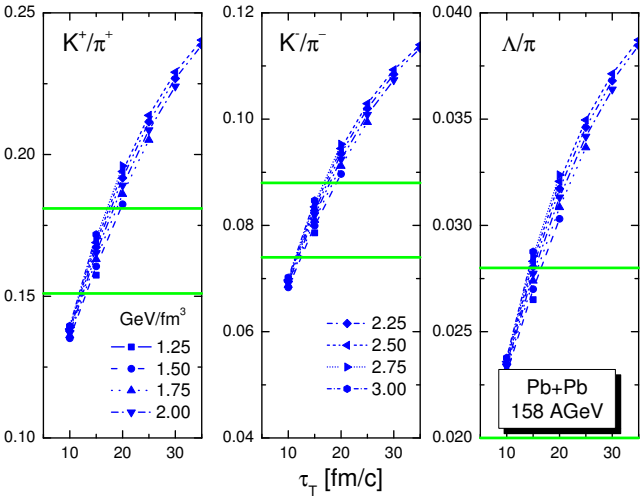}}}
\caption{(colour online) Same as Fig.~\ref{f:AGS} but for Pb+Pb
collisions at projectile energy 158~$A$GeV.
}
\label{f:spo}
\end{figure}

Some selected results are shown in Figs.~\ref{f:AGS}, \ref{f:tri}, and
\ref{f:spo}.
In general, we observe that the resulting ratios mainly depend on
the total lifetime and less so on the initial energy density. It seems
that the data values can be reproduced with rather reasonable
lifetimes, though somewhat longer than one would infer from analysis
of correlation radii based on Bjorken hydrodynamic solution.

\begin{figure}[]
\centerline{\resizebox{0.37\textwidth}{!}{\includegraphics{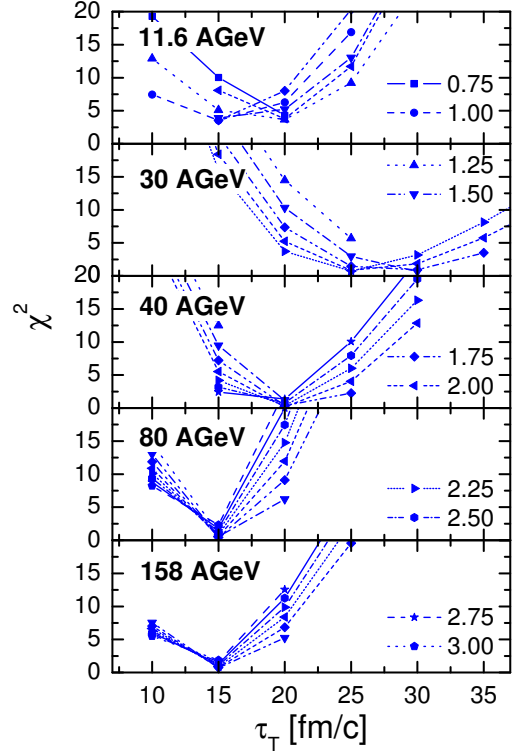}}}
\caption{(colour online) Quantity $\chi^2$ defined in eq.~\eqref{chi}
evaluated for results from all simulations. Plots show the dependence
on total lifetime and different curves correspond to different initial
energy densities.
}
\label{f:chi}
\end{figure}

In order to make our comparison with data more quantitative, we calculate 
for each set of results
\begin{equation}
\label{chi}
\chi^2 = \sum_i \frac{(d_i - t_i)^2}{\sigma_i^2}\, ,
\end{equation}
where $t_i$, $d_i$, and $\sigma_i$ are the calculated and measured values
and experimental error, respectively. For all calculated scenarios
the values of $\chi^2$ are summarised in Fig.~\ref{f:chi}.
We observe that the maximum of $K^+/\pi^+$ ratio of multiplicities
as a function of collision energy 
is basically translated into a maximum lifetime of the fireball
for the best reproduction of data, although such a conclusion is
not statistically significant. One could speculate if it is
connected with long total lifetime due to a soft point in the equation 
of state which would appear close to the phase transition. 

\begin{figure}[t]
\centerline{\resizebox{0.48\textwidth}{!}{\includegraphics{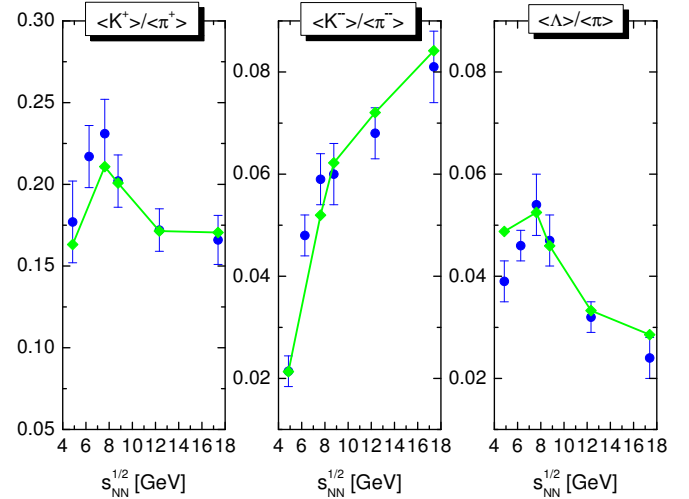}}}
\caption{(colour online) Comparison of selected results of our simulations
with data. Plotted are excitation functions of multiplicity ratios of
$K^+/\pi^+$ (left panel), $K^-/\pi^-$ (middle), and $\Lambda/\pi$ (right).
Data points from \cite{lung}.}
\label{f:comp}
\end{figure}

On the other hand, based on comparison with data one neither can 
exclude that the total lifetime of the fireball does not increase 
and rather decreases when the collision energy grows. In order to demonstrate 
this we choose for each collision energy one set of parameters, 
and show results calculated with those parameters compared to data 
in Fig.~\ref{f:comp}.
The lifetimes and initial energy densities from the used parameter sets
are collected in Table~\ref{t:parms}. 
\begin{table}
\caption{%
Initial energy densities and total lifetimes from parameter sets
which were used in calculations leading to results shown in 
Fig.~\ref{f:comp}. In the lower portion of the table, $T_f$ is the final 
state temperature obtained in our simulations and $T$ the temperature 
from the analysis of chemical freeze-out \cite{becat}.
}
\label{t:parms}
\begin{center}
\begin{tabular}{c|ccccc}
\hline\hline
$E_{\rm beam}$ [$A$GeV] & 11.6 & 30 & 40 & 80 & 158
\\ \hline
$\varepsilon_0$ [GeV/fm$^3$] & 1 & 1.5 & 2 & 2.25 & 2.75
\\
$\tau_T$ [fm/$c$] & 25 & 25 & 20 & 15 & 15
\\
\hline
$T$ [MeV] & 118.1 & 139.0 & 147.6 & 153.7 & 157.8
\\
$T_f$ [MeV] & 114.7 & 134.1 & 143.3 & 149.3 & 153.6
\\ \hline\hline
\end{tabular}
\end{center}
\end{table}

From the table we also see that the final state temperature which we obtain 
does not differ much from the result of chemical freeze-out analysis. Hence, 
we conclude that not only kaon production, but the whole 
calculated chemical composition is correct.


\section{Conclusions}
\label{conc} 

The actual aim in search for the onset of deconfinement is to falsify 
hadronic description of data. This has not been accomplished here. 
Our hadronic kinetic model was able to describe the excitation 
function of the multiplicity ratios $K^+/\pi^+$, $K^-/\pi^-$, and 
$\Lambda/\pi$. It requires, certainly at least for beam energies above
30~$A$GeV, that the lifetime of fireball decreases as the energy of 
collisions grows.

So far, we only checked our model against one type of data. In order 
to verify or falsify it, it will have to be tested on other 
interesting excitation functions mentioned in the Introduction as well 
as the ``standard'' data like, transverse momentum spectra and correlation
radii. Since we make statements about the whole evolution of
the fireball, it seems important to calculate our prediction for 
dilepton spectra which are produced during the whole evolution. 

Finally, so far we have put aside the question of how the evolution which 
we parametrised can result from microscopic structure of the matter.
If the model passes all data tests, this question will have to be addressed.

\section*{Acknowledgements}

BT thanks the organisers of Hot Quarks meeting for creating a
stimulating and friendly atmosphere in Villasimios. He also thanks
OZ Pr\'iroda and European Commission for financial support of his
attendance to this conference. This work has been supported by a
Marie Curie Intra-European fellowship within the 6th European
Community Framework programme (BT) and by the US Department of Energy
under contract No.\ DE-FG02-87ER40328 (EEK).

\end{document}